\documentclass[prl,twocolumn,aps,amsmath,superscriptaddress,showpacs]{revtex4}
\usepackage{graphicx}
\begin{document}
\title{Ordering phenomena in cooling granular mixtures}

\author{C.~Cattuto}
\affiliation{Frontier Research System,
The Institute of Physical and Chemical Research (RIKEN),
Wako-shi, Saitama 351-0198, Japan}
\author{U.~Marini Bettolo Marconi}
\affiliation{Dipartimento di Fisica, Universit\`a di Camerino and
Istituto Nazionale di Fisica della Materia,
Via Madonna delle Carceri, 62032 Camerino, Italy}

\begin{abstract}
We report two phenomena, induced by dynamical correlations, 
that occur during the free cooling
of a two-dimensional mixture of inelastic hard disks.
First, we show that, due to the onset of velocity correlations,
the ratio of the kinetic energies
associated with the two species changes from the value corresponding to 
the homogeneous cooling state to a value approximately given by the mass ratio
$m_1/m_2$ of the two species. Second, we report a novel segregation effect
that occurs in the late stage of cooling, where interconnected domains appear.
Spectral analysis of the composition field reveals the emergence of a
growing characteristic length.
\end{abstract}

\maketitle

\begin{figure}
\includegraphics[clip=true,width=\columnwidth, keepaspectratio]{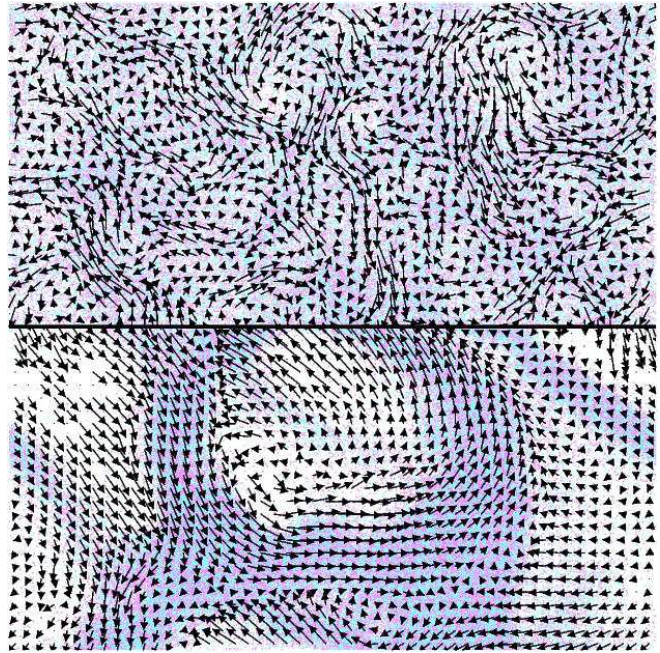}
\caption{\label{snapshot}
Status of the system at points A (upper panel) and B (lower panel)
of fig.~\ref{due}. The instantaneous velocity field
is superimposed. Early vortices can be seen in the upper panel,
with particles still (approximately) uniformly distributed.
In the lower panel, dense clustering has appeared
and the typical size of vortices has significantly increased.
Strong correlations in the velocity field are visible.
}
\end{figure}

A gas of inelastic hard spheres (IHS),
due to its relative simplicity, represents a standard reference model
for fluidized granular materials \cite{general}.
Many of its properties are well understood.
In particular, a great deal of attention has been devoted to the study
of the cooling process that occurs when an assembly of grains, initially 
in motion, evolves in the absence of any external energy feed.
In this case, while the asymptotic state
(all particles at rest) is trivial, the dynamics of the cooling process
displays several interesting features.
Three stages can be identified:

i) After an initial transient, during which the velocity
distribution is not necessarily Maxwellian~\cite{gold1}, the system enters the
homogeneous cooling state (HCS),
where the average kinetic 
energy decreases according to Haff's law \cite{Haff}, $K(t)=K(0)/(1+t/t_0)^2$.
Such a state is characterized by
a uniform density and a lack of strong correlations among the 
velocities \cite{Poschel}. 

ii) An intermediate regime, wherein the system remains homogeneous,
but the velocity field develops vortices (fig.~\ref{snapshot}).
These vortices result from the parallelization of velocities
induced by inelasticity.
Due to the progressive build up of velocity correlations,
the kinetic energy decays as $t^{-d/2}$ \cite{Zanetti,Brito,Brey}.
 
iii)  Goldhirsch
and Zanetti~\cite{Zanetti} suggested that vortices represent
the physical mechanism leading to the third stage, 
wherein the density field itself
loses its homogeneity and dense clusters appear,
surrounded by rarefied regions.
At this stage neither the velocity nor the temperature fields are homogeneous
anymore (see fig.~\ref{snapshot}).

Most of the existing information on the cooling properties 
of granular matter concern one-component systems.
A few investigations on
binary mixtures \cite{dufty1,noantri} focused on the HCS only. 
Garz\'o and Dufty  \cite{dufty1} showed that 
for a binary mixture the average kinetic energies per particle
-- i.e. the kinetic temperatures $K_1$ and $K_2$
of the two species -- are different, and their ratio $\gamma=K_1/K_2$
remains constant in the HCS. 

In this Letter we explore the cooling dynamics of a mixture of inelastic
hard disks across the previously listed dynamical regimes
by using an event-driven molecular dynamics simulation
and a recently introduced lattice model \cite{Baldassa}.
We show that during the second stage of cooling -- at the onset of local
velocity correlations -- the ratio of the kinetic temperatures of the two
species sharply departs from the nearly steady value attained during
the homogeneous cooling stage, and eventually approaches
an asymptotic value that only depends on the mass ratio of the two species.
In the last stage we find evidence for segregation of the species,
which is revealed both by local indicators and by spectral analysis
of the composition field.

For our model we consider a two-dimensional (2D) system containing
$N_1 = 5 \times 10^4$ and $N_2 = N_1 $ disks of equal radius $R$
and masses $m_1$ and $m_2$, respectively. 
Initially, the discs are uniformly distributed in a square domain of side
$L = 10^3 R$, with periodic boundary conditions, and their velocities
are set by sampling Maxwell distributions corresponding
to the same temperature. In the IHS model the collisions
are instantaneous, conserve total momentum and dissipate a fraction
of the kinetic energy. The velocity ${\bf v_j'}$ of a particle $j$
after collision is
\begin{equation}
{\bf v_j'}={\bf v_j} +
\mu_{ij} \, [ 1+r(v_n) ] \,
v_n \, {\bf n} \,\, , 
\label{rule}
\end{equation}
where ${\bf n}=({\bf r}_i - {\bf r}_j) / |{\bf r}_i - {\bf r}_j|$
is the unit vector joining the centers ${\bf r_i}$ and $\bf{r_j}$
of the disks at contact, $v_n = {\bf(v_i-v_j)} \cdot {\bf n}$
is the relative normal velocity of the disks before collision,
$\mu_{ij} = m_i / (m_i+m_j)$,
and $r(v_n)$ is a velocity-dependent coefficient of restitution.
Inelastic collapse is avoided by choosing a $r(v_n)$ such that,
for relative normal velocities falling below a threshold $v_0$,
$r$ approaches unity when $v_n \rightarrow 0$,
according to the power law given in ref. \cite{Bizon}.
Different cutoff velocities were simulated in order
to ascertain the robustness of the investigated properties.
An efficient event-driven simulation allowed us to 
probe the late stages of the cooling process \cite{Paolotti}.
  
\begin{figure}
\includegraphics[clip=true,width=\columnwidth, keepaspectratio]{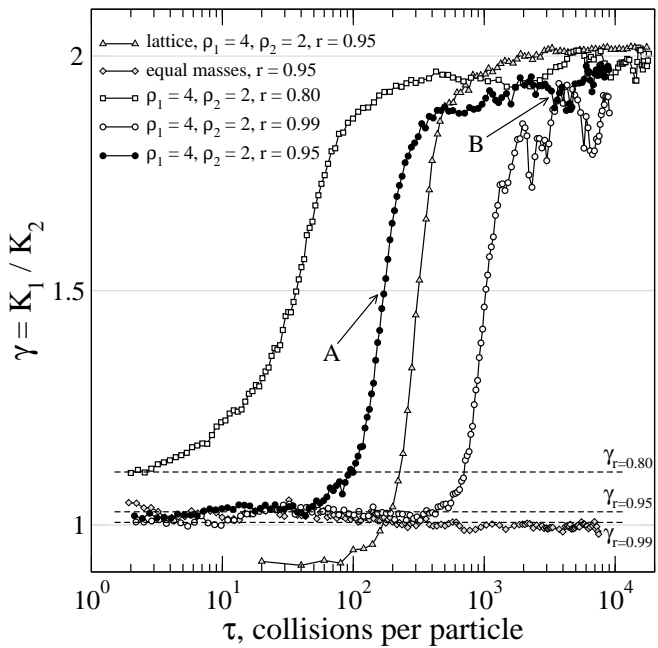}
\caption{\label{due}
Ratio of the kinetic energies of the two species for different 
values of the coefficient of restitution. The crossover from $K_1/K_2 \sim 1$
to $K_1/K_2 \sim m_1/m_2$ is clearly visible.
For comparison, we also show the corresponding result for the lattice model.
The status of the system at points
A and B is shown in fig.~\ref{snapshot}.
}
\end{figure}

During the initial homogeneous cooling phase, the behavior of the mixture is
well described by the Boltzmann-Enskog (BE) transport equation \cite{dufty1}.
Theoretical values for the ratio $\gamma$ can be computed within this
framework, and their values for the chosen set of microscopic parameters
are shown in fig.~\ref{due} (dashed lines). These theoretical values agree
with the initial plateau observed in the temporal evolution of $\gamma$.
However since the BE approach, by construction,
neglects velocity correlations, one can expect that the intermediate
cooling regime cannot be properly described within its framework.
Indeed, hydrodynamics predicts that the HCS is unstable with respect
to long-wavelength fluctuations of the velocity field. 
Its repercussions on the behavior of a mixture
have not been explored to date.
We verified that $K_1(t)$ and $K_2(t)$
exhibited -- separately -- the $t^{-2}$ and $t^{-1}$
scaling behaviors that are commonly observed in
one-component systems.

Fig.~\ref{due} shows that $\gamma = K_1 / K_2$ changes rather sharply from
the initial plateau value of the HCS to a higher plateau value.
The number of collisions at which the crossover occurs is proportional
to $(1-r^2)^{-1}$ and diverges when the system approaches perfect elasticity.
As we mentioned above, the position of the lower plateau is well accounted
for by the BE theory. From fig.~\ref{due} one can also
see that the asymptotic, higher plateau corresponds to $K_1/K_2=m_1/m_2$.
This correspondence was confirmed for several values of the density ratio
of the two species.
Whereas in the HCS the inelasticity determines only a small departure
from equipartition, in the intermediate and late regime such a breakdown
becomes more severe: the specific asymptotic value of $\gamma$ means
that particles with different masses eventually share the same velocity 
distributions, i.e. ``equipartition'' of velocities eventually sets in.
Although not previously reported, the phenomenon has a simple explanation:
due to the inelasticity of collisions, the velocities of the particles
tend to become more and more parallel.
Since momentum conservation does not allow
the velocities to select a unique direction,
the velocity field develops many vortices (see fig.~\ref{snapshot}),
i.e. a situation of local alignment.
The occurrence of a collision depends only on
the relative velocities of particles,
and not on their masses, coefficients of restitution or other properties.
Therefore, the velocity of a given particle tends to be locked to
the velocity of neighbors, regardless of their species.
The fact that the ratio $K_1/K_2$ eventually attains the value $m_1/m_2$
is a striking manifestation of the long range velocity
correlations that develop in a granular fluid.
We checked the above argument against a different case, wherein
densities are kept unchanged and the masses of the two species are made
equal by choosing different radii for them. The result is shown
in fig.~\ref{due} (gray diamonds). In this case the value of $\gamma$
stays close to $m_1/m_2=1$.
Also, the dependence on other parameters (such as different restitution
coefficients, geometrical factors, tangential friction modeled as in
ref. \cite{Paolotti}) was found to be rather weak.
Of course, real granular materials are subject to more complex
dissipative forces. Our tests, however, show that $\gamma$ asymptotically
depends mainly on $m_1/m_2$, although the detailed crossover dynamics
reflects the specific parameters.

It is interesting to observe that the previously presented scenario
is reproduced by a very simple extension to mixtures
of the inelastic lattice model of ref. \cite{Baldassa}. The mixture consists
of $N_1$ and $N_2$ ``particles'' with the same physical properties as above,
sitting randomly on the $M=N_1+N_2$ nodes of a 
2D triangular lattice. Every particle is characterized by a
2D vector velocity $\bf{v}_i$, but its position stays fixed.
A dynamics is induced by randomly choosing two neighboring particles
and updating their velocities according to the rule of eq. \ref{rule}.
The free streaming component of the dynamics is neglected,
but velocity correlations are properly taken into account.
After an exponential decay, the energy decreases as $K \sim \tau^{-d/2}$,
as usual. Correlations become more and more important as time elapses,
showing the presence of a growing length scale $L(\tau)\sim \tau^{1/2}$
associated with the average diameter of the vortices \cite{Baldassa}.
As shown in fig.~\ref{due} (gray triangles) the lattice mixture model
can reproduce the asymptotic value of $\gamma$.
To demonstrate that velocity correlations are responsible for the specific
value of $\gamma$, we shuffled the velocity of each particle right after
collision, by exchanging its velocity with that of a randomly chosen
particle. This suppresses completely the $\tau^{-d/2}$ energy decay,
the vortex instability, and the crossover in $\gamma$.

\begin{figure}
\includegraphics[clip=true,width=\columnwidth, keepaspectratio]{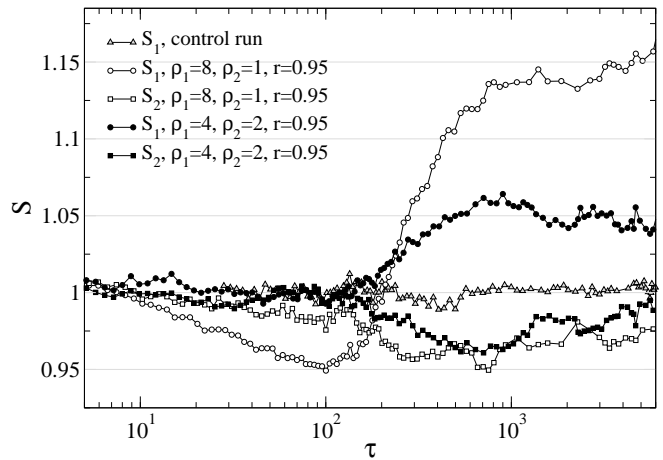}
\caption{\label{segrega}
Local segregation: the temporal evolution of the ratios $S_1$
and $S_2$ of integrated correlation functions (see text)
is shown for $m_1/m_2 = 2$ and $m_1/m_2 = 8$.
To assess noise levels, a control run with $m_1 = m_2$ is also shown.
} 
\end{figure}

\begin{figure}[t]
\includegraphics[clip=true,width=\columnwidth, keepaspectratio]{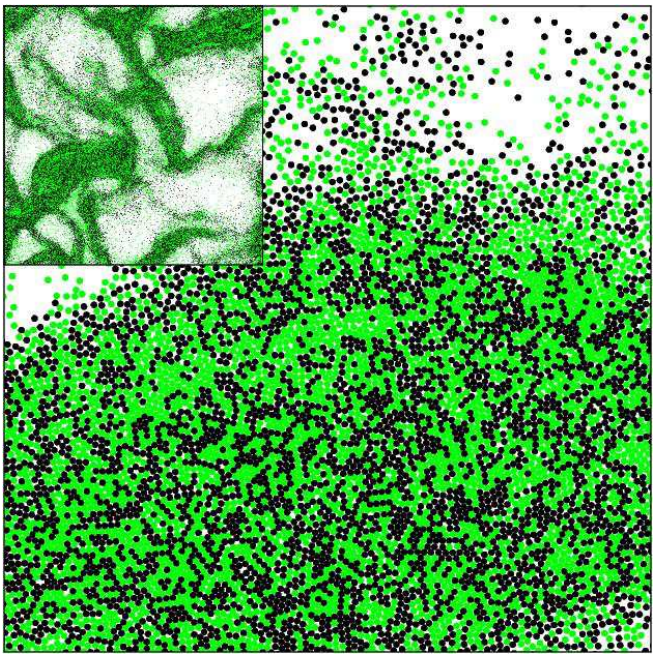}
\caption{\label{spinodale}
Composition pattern within a cluster after $\tau=1300$ collisions
per particle, for $r = 0.95$ and $m_1 / m_2 = 8$. Black discs correspond
to heavier particles. The radius of the particles is $R=0.1$ cm.
In inset: global state of the system.
}
\end{figure}

In the third stage of cooling, density fluctuations lead to cluster formation,
like in mono-disperse systems. In addition to that (see fig.~\ref{spinodale}),
we observe a novel and appreciable segregation of species,
that we quantify by studying
the following ratios $S_1$ and $S_2$ of integrated (partial)
pair distribution functions:
$ S_{\alpha}  =  \int_0^{D} \! g_{\alpha\alpha}(r) \, r \, dr \, / \, 
\int_0^{D} g_{\alpha\beta}(r) \, r \, dr$, 
where $g_{\alpha\beta}(r)$ ($\alpha, \beta = 1,2$) are the partial
pair correlation functions
and $D$ is a cutoff distance, here chosen equal to $5R$.
Fig.~\ref{segrega} shows the time evolution of $S_1$ and $S_2$
for different choices of the mass ratio. For values of $\tau$
corresponding to the transition in $\gamma$ (refer to fig.~\ref{due}),
local segregation appears: heavy particles tend to group together
($S_1>1$) and the effect increases with the mass ratio.
A simple heuristic argument can explain such a finding.
The average change of kinetic energy
$\langle \Delta E_\alpha\rangle_\beta$
of a particle of species $\alpha$ colliding with a particle of species $\beta$
can be expressed as a function of the local kinetic temperatures
\cite{footnote}
$T_1$ and $T_2$ as \cite{Talbot}
\begin{eqnarray}
\langle \Delta
E_\alpha\rangle_\beta & = & \frac{m_{\alpha}m_{\beta}}{(m_{\alpha}+m_{\beta})}
\frac{(1+r)}{2\sqrt{2}} \cdot \nonumber \\
& & \biggl[ (1+r)\frac{T_\beta m_\alpha+T_\alpha m_\beta}{m_\alpha(m_\mu+m_\beta)}- 2\frac{T_\alpha}{m_\alpha} \biggr] \, .
\label{deltaE}
\end{eqnarray}
One sees that particles with a smaller mass gain energy
($\langle \Delta E_2\rangle_1>0$) at the expenses of the heavy particles,
provided that $T_1/T_2 > ((1-r)m_1/m_2+2)/(1+r) > 1$.
Indeed, our numerics confirm that this condition is satisfied for our system.
For a perfectly elastic system there is no energy transfer, since $T_1=T_2$.
In the dissipative case, instead, since the ratio of the kinetic
temperatures $T_1/T_2>1$, light particles -- on average -- gain kinetic energy
when colliding with heavy particles.
This results in an effective repulsion
of the light particles by the heavy ones.
The above scenario is also consistent with the asymmetry between species
observed in the temporal evolution of the indicators $S_{\alpha,\beta}$
of local segregation.

\begin{figure}[b!]
\includegraphics[clip=true,width=\columnwidth, keepaspectratio]{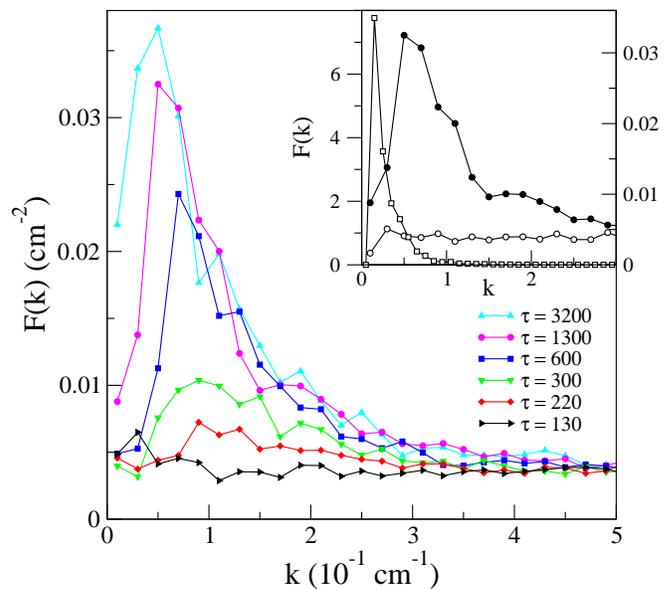}
\caption{\label{cahnhilliard}
Circularly averaged composition structure factor, $F(k,t)$, at different
times. With time, the peak grows higher and moves towards smaller
wavevectors. The inset shows how the peak for $\tau = 1300$ (solid circles,
right vertical axis) is suppressed on reshuffling the particle species
(empty circles, right axis). The structure factor of the total density field
(empty squares, left axis) has a peak at a much lower wavevector.
For these curves, $r = 0.95$ and $m_1 / m_2 = 8$, as in fig.~\ref{spinodale}.
}
\end{figure}

Such a microscopic mechanism driving segregation has also global effects.
Let us consider the long-wavelength properties of the composition
field $\phi({\bf r},t) \equiv n_1({\bf r},t)-n_2({\bf r},t)$,
the difference of the local number densities of
the two species, whose dynamics is slow due to its conservative nature.
The analysis of the structure factor
$F({\bf k},t)=\hat \phi({\bf k},t)\hat \phi^{\ast}({\bf k},t)$
(where $\hat \phi({\bf k},t)$ is the Fourier transform of $\phi({\bf r},t)$)
reveals a remarkable scenario:
at the onset of clustering the uniform composition state becomes unstable
and evolves into an interconnected pattern of domains
with different composition, as shown in fig.~\ref{spinodale}.
Long-wavelength fluctuations of $\phi({\bf r},t)$ are amplified and --
contrary to the usual diffusion processes -- particles
of different species tend to segregate. The temporal evolution of the above
(circularly averaged) structure factor is reported in fig.~\ref{cahnhilliard}.
In the homogeneous stage of cooling, the structure factor is featureless
(right triangles). When clustering appears, a broad peak develops (diamonds)
and with time it grows higher, while shifting towards low wavevectors.
Reshuffling the particle species -- while keeping their positions unchanged --
suppresses the peak altogether,
as shown in the inset of fig.~\ref{cahnhilliard}.
This proves that the above spectral properties are truly a manifestation
of progressive, long-wavelength ordering of the two species. Moreover,
the peak in $F(k,t)$ is unrelated to the short wavevector peak that
-- because of clustering -- is expected in the structure factor of the total
density field (see fig.~\ref{cahnhilliard}, inset): the composition field
and the density field have very different characteristic lengths.

In conclusion,
we found that the strong velocity correlations induced by
inelasticity result in a ratio of kinetic energies
$K_1/K_2 \simeq m_1/m_2$ that is not accounted for by the BE theory,
which underestimates the effect of dynamical correlations. This can be
regarded as a manifestation of the noise reduction effect \cite{Ernst}
of granular systems, where a suitable collective arrangement of the velocities
minimizes dissipation. In addition to that,
in the clustered stage, segregation builds up according to
a phenomenology which bears similarities to the spinodal decomposition
occurring in metallic alloys, where a growing characteristic
length appears in the composition field.
C.~C. gratefully acknowledges support from the Japan Society for the Promotion of Science (JSPS).

\begin{thebibliography}{99}
\bibitem{footnote}
Local kinetic temperatures are computed by subtracting the local
macroscopic drift velocity of particles prior to computing their
kinetic energies.
\end{thebibliography}
\end{document}